 \documentclass[twocolumn]{aastex6}
 \usepackage{lineno}

\newcommand{\dechms}[4]{$#1^{\rm h}#2^{\rm m}#3\mbox{$^{\rm s}\mskip-7.6mu.\,$}#4$}

\newcommand{\decdms}[4]{$+#1^{\circ}#2'#3\mbox{$''\mskip-7.6mu.\,$}#4$}

\begin{document}

\newcommand{\be}{\begin{equation}}
\newcommand{\ee}{\end{equation}}
\newcommand{\eint}{{ E_{\rm int} }}
\newcommand{\emix}{{ E_{\rm mix} }}
\newcommand{\excl}{{ E_{\rm excl} }}
\newcommand{\eself}{{ E_{\rm self} }} 
\newcommand{\ewig}{{ {\widetilde E} }}
\newcommand{\zhat}{{ {\hat z} }}
\newcommand{\lmech}{{ L_{\rm mech} }}
\newcommand{\fmean}{{ f_{\rm mean} }}  
\newcommand{\porb}{{ P_{\rm orb} }}
\newcommand{\ergsec}{{ {\rm erg}\,{\rm s}^{-1} }}
\newcommand{\asemi}{{ a_{\rm o} }} 
\newcommand{\pmag}{{ P_{\rm mag} }} 
\newcommand{\bint}{{ B_{\rm int} }} 
\newcommand{\numax}{{ \nu_{\rm max} }} 
\newcommand{\msun}{{ M_\odot }} 
\newcommand{\kms}{{{\rm km\, s^{-1}}}}  
\newcommand{\yr}{{{\rm yr^{-1}}}} 
\newcommand{\AU}{{{\rm AU}}} 
\newcommand{\mfl}[1]{{\color{blue} \bf #1}}


\title{A disk-wind driving the rotating molecular outflow in CB 26}



\author{J. A. L\'opez-V\'azquez\altaffilmark{1,2}, Luis A. Zapata\altaffilmark{2}, and Chin-Fei Lee\altaffilmark{1}} 


\altaffiltext{1}{Academia Sinica Institute of Astronomy and Astrophysics, No. 1, Sec. 4, Roosevelt Road, Taipei 10617, Taiwan}
\altaffiltext{2}{Instituto de Radioastronom\'ia y Astrof\'isica, Universidad Nacional Aut\'onoma de M\'exico, 
Apartado Postal 3-72, 58089 Morelia, Michoac\'an, M\'exico}


\begin{abstract}
We present $^{12}$CO (J=2--1) sensitive molecular line  and 1.3 mm continuum observations made 
with the Submillimeter Array (SMA) of the bipolar outflow associated with the young star located in 
the Bok globule known as CB 26.  The SMA observations were carried out in its extended configuration
allowing us to study the kinematics and structure of the outflow with about 1$''$  or 140 au resolution.  
We find that the dusty and edge-on circumstellar disk related with the outflow has an projected spatial 
(deconvolved) size of 196$\pm$31 au$\times$42$\pm$29 au with
a total (gas+dust) mass of 0.031$\pm$0.015 M$_\odot$. We estimated a dynamical mass for the central 
object of 0.66$\pm$0.03 M$_\odot$, and the mass of the molecular outflow of $5\pm1.5\times10^{-5}$ M$_{\odot}$.  
All these values are consistent with recent estimations.
The observations confirm that the outflow rotation has a similar orientation to that of the edge-on disk.
For the outflow, we find that the following quantities: the rotation velocity ($\sim$1--3 km s$^{-1}$), 
the specific angular momentum ($\sim$200--700 au km s$^{-1}$),  and the launching radius ($\sim$15--35 au), 
decrease with the height above mid--plane, as observed in other molecular rotating outflows. 
The radius ($\sim$180--280 au), and expansion velocity ($\sim$2--4 km s$^{-1}$) also increase with the height 
above the disk mid–plane for $z<0$ au, however, for $z>0$ au these quantities do not present this behavior.
Estimations for the outflow linear momentum rate, the outflow angular momentum rate, and the accretion luminosity seem to be
well explained by a disk-wind present in CB26.

\end{abstract}

\section{Introduction}
\label{sec:introduction}

Rotating molecular outflows and protostellar jets are present in the early phases of the star formation process.
They are thought to play an essential role in reducing 
the angular momentum from the disk-protostellar system (e.g., \citealt{Blanford_1982} and \citealt{Machida_2014}) since it is proposed that 
these objects are ejected directly from the accretion disks (e.g., \citealt{Pudritz_1986} and \citealt{Bai_2016}). 
The molecular outflows in addition limit the mass of the star-disk system \citep{Shu_1993}, and can induce changes in the 
chemical composition of their host cloud (e.g., \citealt{Bachiller_1996}), it is because some authors have proposed that the 
molecular outflows are a mixture between the entrained material of the molecular 
cloud and a stellar or disk wind (e.g., \citealt{Shu_1991} and  \citealt{JALV_2019}).
Moreover, the protostellar jets could be helpful in the determination of the initial environmental properties related with planetary formation, 
given that these objects are able to remove the angular momentum of the material in the disk, allowing a continuous accretion towards the star. 
This is possible because planet formation begins much earlier than previously expected \citep{Ray_2021}.
 
Independent of their origin, it is expected that the molecular outflows present rotation signatures because they should inherit 
an angular momentum component from the accretion disk, or the parent cloud, or a combination of both.  
Recent studies have reported the rotation signatures in several sources, 
the molecular outflows with rotation are: CB 26 \citep{Launhardt_2009}, Ori-S6 \citep{Zapata_2010}, HH 797 \citep{Pech_2012}, 
DG Tau B (\citealt{Zapata_2015} and \citealt{DeValon_2020}), TMC1A \citep{Bjerkeli_2016}, Orion Source I (\citealt{Hirota_2017} 
and \citealt{JALV_2020}), HH 212 (\citealt{Lee_2018}) HH 30 \citep{Louvet_2018}, NGC 1333 IRAS 4C \citep{Zhang_2018}, 
and HD 163296 \citep{Booth_2021}. While, the protostellar jets with rotation are: HH 211 (\citealt{Lee_2009}), HH 212 (\citealt{Lee_2017}), 
and OMC 2/FIR 6b (\citealt{Matsushita_2021}).

Located at 10$^\circ$ north of the Taurus Auriga dark cloud at a distance of 140$\pm$20 pc \citep{Launhardt_2009}, 
CB 26 is a Class I young stellar object \citep{Stecklum_2004} with a dynamical mass of the central star of ${\rm M}_*=0.55\pm0.1$ M$_\odot$, 
and an estimated age of about 1 Myr \citep{Zhang_2021}. 
The central star has a luminosity ${\rm L}_*\ge$0.5 L$_\odot$ \citep{Stecklum_2004}. The molecular outflow associated with CB 26 
has an outward velocity of $\sim$10-12 km s$^{-1}$ along the southwest-northeast direction (\citealt{Pety_2006} and \citealt{Launhardt_2009}), with a diameter of 2000 AU \citep{Launhardt_2009}.

We present new $^{12}$CO (J=2--1) molecular line observations, made with the Submillimeter Array (SMA) of the 
molecular outflow associated with the young star CB 26. This paper is organized in the following way: Section \ref{sec:observations} 
details the observations. In Section \ref{sec:results} we present the results. In Section \ref{sec:discussion} we discuss our results. 
Finally, the conclusions are presented in Section \ref{sec:Conclusions}.

\begin{figure}[t!]
\centering
\includegraphics[scale=0.53]{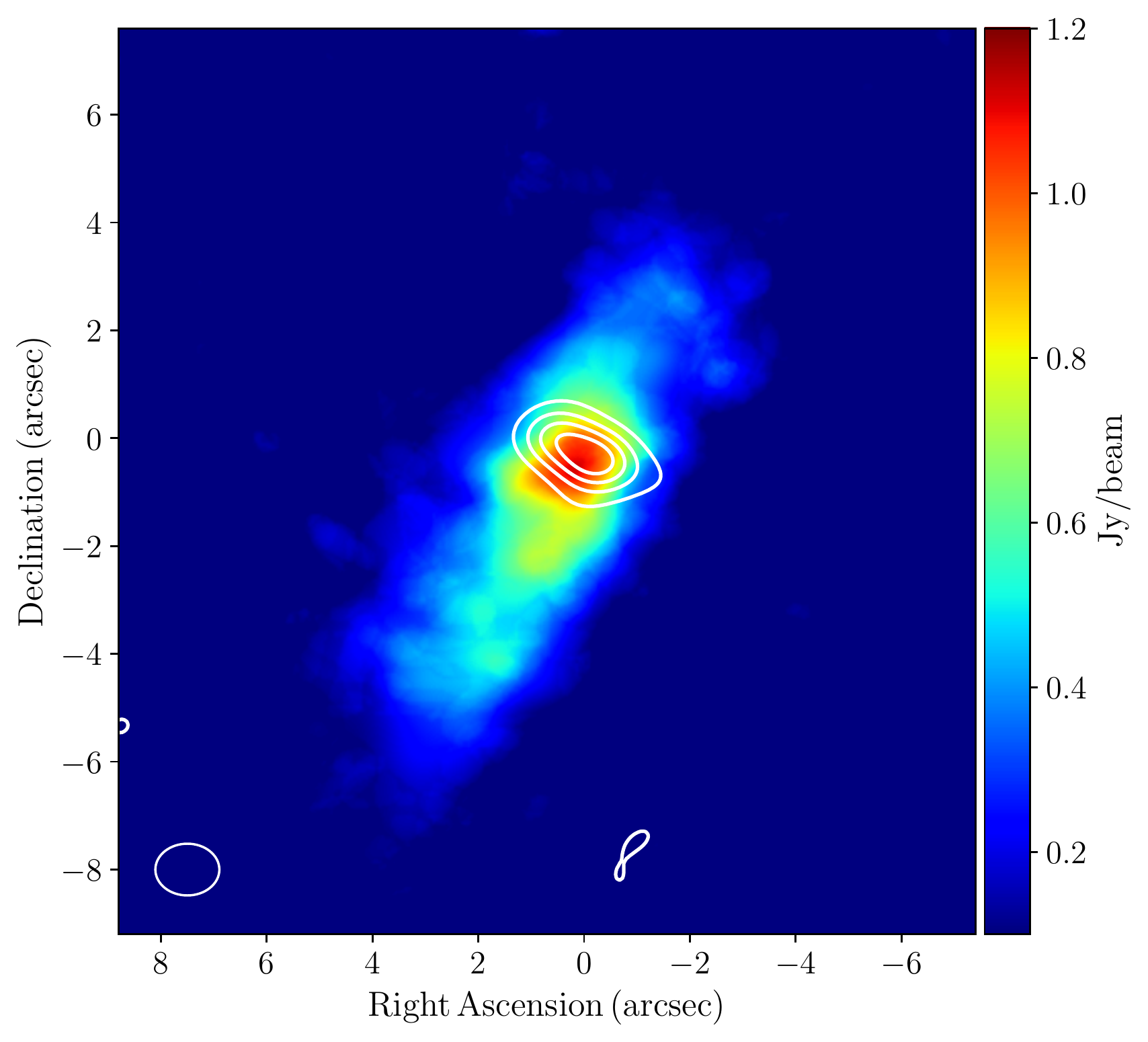}
\caption{Moment zero of the $^{12}$ CO (J=2--1)  emission line from the molecular outflow. 
The color scale bar on the right side shows the intensity in Jy/beam. The synthesized beam of the image is shown in the lower left corner. 
The white contours show the continuum emission from the disk and are the 16.5 mJy, 33.0 mJy, 49.5 mJy, and 66.0 mJy.}
\label{fig:mom0}
\end{figure}

\begin{figure}[t!]
\includegraphics[scale=0.525]{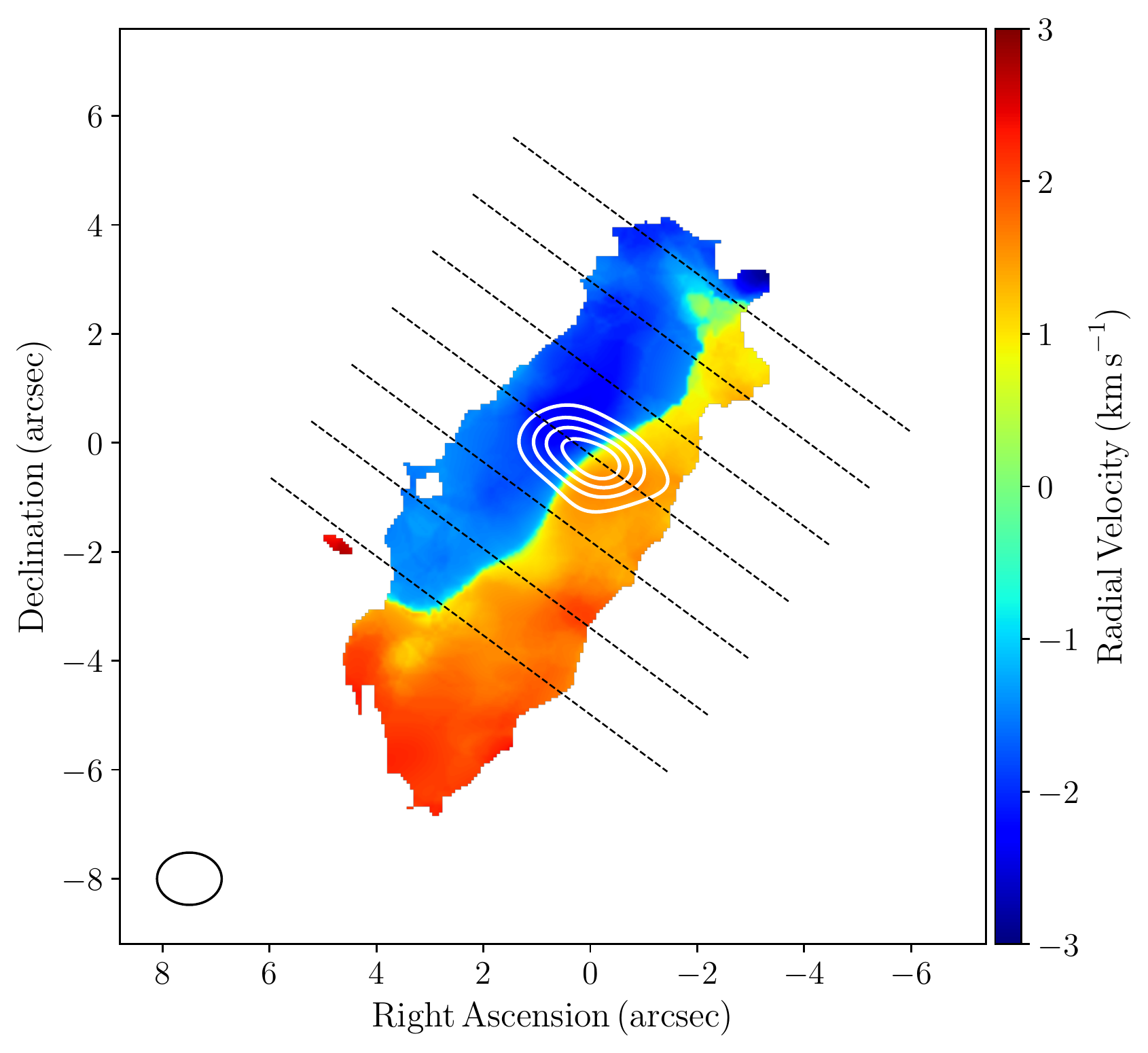}
\caption{First moment or the intensity weighted velocity of the $^{12}$CO (J=2--1)  emission line from the molecular outflow. 
The diagonal dashed lines indicate the positions where the position-velocity diagrams were made (see Figure \ref{fig:pvz}). 
The color scale bar on the right side shows the $V_{\rm LSR}$ in km s$^{-1}$. The synthesized beam of the image is shown in the lower left corner. 
The white contours show the continuum emission from the disk and are the 16.5 mJy, 33.0 mJy, 49.5 mJy, and 66.0 mJy.}
\label{fig:mom1}
\end{figure}

\label{subsec:structure}
\begin{figure*}[t!]
\includegraphics[scale=0.53]{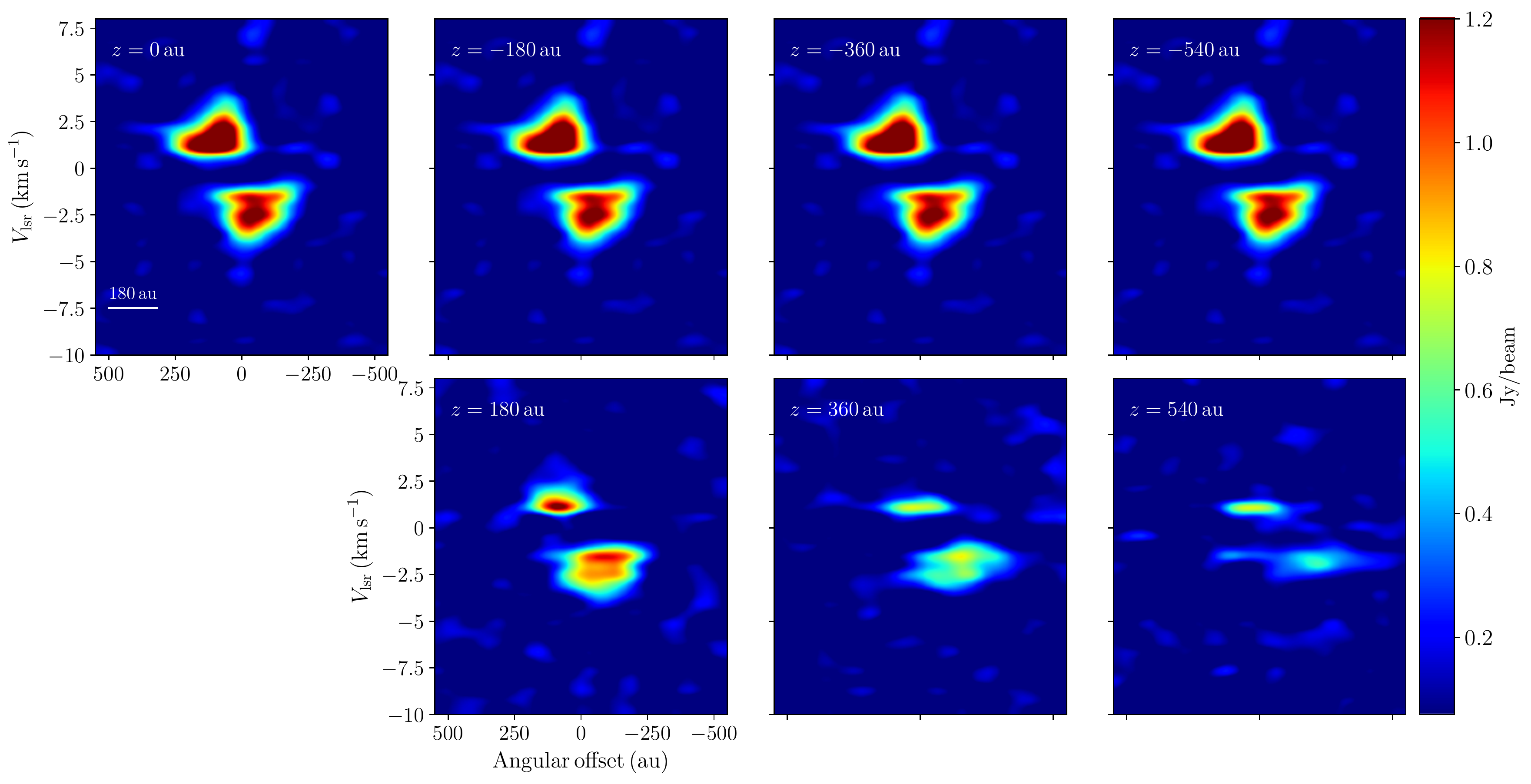}
\caption{Position-velocity diagrams parallel to the disk mid-plane from the emission of the 
$^{12}$CO (J=2--1) transition at different heights from $z=-540$ au to $z=540$ au with an interval of 180 au. 
The vertical axes are the line of sight velocity with respect to the LSR velocity and the horizontal axes are the perpendicular distances with respect to the outflow axis. 
The color scale bar on the right side shows the intensity in Jy/Beam. The white bars indicate the angular resolution (140 au or 1'') and velocity resolution (0.18 $\kms$) used for the position--velocity cuts.}
\label{fig:pvz}
\end{figure*}

\section{Observations}
\label{sec:observations}

The Submillimeter Array\footnote{The Submillimeter Array is a joint project between the Smithsonian Astrophysical 
Observatory and the Academia Sinica Institute of Astronomy and Astrophysics, and is funded by the Smithsonian 
Institution and the Academia Sinica.} observations of CB26 were carried out on October 4 2013 when the array 
was in its extended configuration. The baselines in this configuration ranged  from 15 to 148 k$\lambda$. 
The phase center in the sky was centered to $\alpha_{J2000.0}$ = \dechms{04}{59}{50}{7}, 
and $\delta_{J2000.0}$ = \decdms{52}{04}{43}{6} with a total integration time on source of 240 min. 
The Full Width Half Maximum (FWHM) of the primary beam is 57$''$ at this frequency, so that the bulk of the molecular 
and dusty material associated with CB26 is well-covered, see \citet{Launhardt_2009}.

Two frequency bands were observed simultaneously centered at 230.457 GHz (upper sideband) and 220.457 GHz (lower sideband). 
The SMA-SWARM correlator was configured to cover 2 GHz bandwidth in both bands.  The native spectral resolution is 140 kHz per 
channel across the entire spectral band. This provided a velocity resolution of about 0.18 km s$^{-1}$. This spectral resolution is bit smaller to that obtained
in \citet{Launhardt_2009}, 0.25 km s$^{-1}$. This very high spectral resolution allowed a reliable study of the CO kinematics in CB26 outflow.

With an average system temperature of about 150 K and with opacity at 225 GHz around 0.12, the observations were made in optimal
conditions. Observations of Neptune and Uranus served for the flux density calibration. The gain calibrators were the quasars 3C 111 and J0533$+$483, 
while 3C454.3 was used for bandpass calibration. The estimated uncertainty in the flux scale, what is based in SMA monitoring, is in a window between 15\% to 20\%.

The IDL superset MIR developed for the Owens Valley Radio Observatory and adapted for the 
SMA\footnote{https://lweb.cfa.harvard.edu/$\sim$cqi/mircook.html} was used for the calibration of the SMA data. 
The calibrated data were then imaged and analyzed in a standard manner using Common Astronomy Software Applications (CASA) 
package. We also used some routines in Python to image the data \citep{ast2013}.  A 1.36 mm continuum image was obtained by 
 averaging line-free channels in the lower sideband with a bandwidth of about 8 GHz. For this image, we used a robust factor of zero in order to obtain 
 an optimal compromise between sensitivity and angular resolution. The continuum image {\it rms}-noise is 
 2.2 mJy beam$^{-1}$ at an angular resolution of 1.1$''$ $\times$ 0.9$''$ with a PA (Position Angle) of $-$85.8$^\circ$. For the line image, we obtained
 a {\it rms}-noise of 90 mJy beam$^{-1}$ km s$^{-1}$ at an angular resolution of 1.2$''$ $\times$ 0.96$''$ with a PA of $-$85.5$^\circ$.
 For the line image, we used a robust factor of 2.0 in order to obtain a better sensitivity. 
 The resulting SMA rms in the channels maps is higher than that obtained in \citet{Launhardt_2009}, 
 which is 20 mJy Beam$^{-1}$ at an effective beam size of 1.47$^{\prime \prime}$.

\section{Results}
\label{sec:results}

Figure \ref{fig:mom0} shows the moment zero map of the $^{12}$CO (J=2--1) emission line overlaid in white contours of the 1.36 mm continuum map. 
This continuum emission is tracing the edge-on disk surrounding the young source CB26  
(e.g., \citealt{Stecklum_2004}; \citealt{Sauter_2009}; \citealt{Akimkin_2012}). The $^{12}$CO (J=2--1)  
 emission extends further out than the continuum emission along the southeast-northwest. We note that the extent of the 
molecular outflow is around $\sim$1600 au, and the width is $\sim$600 au. For the continuum emission, a Gaussian fit to the CB26 disk
resulted in a deconvolved size of 196$\pm$31 au $\times$ 42$\pm$29 au (1.4$''$ $\pm$ 0.1$''$ $\times$ 0.3$''$ $\pm$ 0.2$''$) with position angle of 59$^\circ$ $\pm$ 3$^\circ$,
and an integrated flux of 161 $\pm$ 8 mJy together with a peak flux of 82 mJy Beam$^{-1}$. 
The phase center of CB26 disk is at $\alpha_{J2000.0}$ = \dechms{04}{59}{50}{742}, and $\delta_{J2000.0}$ = \decdms{52}{04}{43}{49}. The integrated flux is indeed lower
to that estimated by \citet{Launhardt_2009}, which is 190 mJy. This could be explained as we are not recovering all the extended emission
given that the effective beam size from \citet{Launhardt_2009} is larger, 1.47$''$. 
However, if we consider that the estimated uncertainty in the flux scale is between 15\% to 20\%, 
our integrated flux could increase to 193$\pm$10 mJy, a value is similar to the one estimated by them.

The first moment or the intensity weighted velocity map of the $^{12}$CO (J=2--1)  emission line is presented 
in Figure \ref{fig:mom1}. We note that the east side of the molecular outflow presents blueshifted velocities, 
while the west side presents redshifted velocities. This difference of the velocities is proposed  as rotation
 around the outflow axis \citep{Launhardt_2009}. The inclination angle of this source respect to the plane of the sky  is 
 $i=5^\circ\pm 4^{\circ}$ \citep{Launhardt_2009}, therefore, the lower edge of the outflow has an excess redshifted velocity.
 Consequently, the putative rotating outflow in CB26 is an excellent object to study the kinematics and nature of the flow. 

Figure \ref{fig:pvz} presents the position-velocity diagrams of the emission from the molecular line of $^{12}$CO (J=2--1). 
The different panels of this figure correspond to parallel cuts at different heights above the disk mid-plane, these cuts were 
made from $z$=--540 au to $z$=540 au with intervals of 180 au (these cuts correspond to the dashed lines in Figure \ref{fig:mom1}). 
We note that all position-velocity diagrams present signatures of rotation, in a range of $\sim$1--3 km s$^{-1}$. 
The position-velocity diagrams presented in this figure does not present a hollow structure, this might be explained as 
we do not have enough angular and spectral resolution to resolve angularly the outflow. This hollow structure is 
shown in other molecular outflows, e.g., Orion Source I (\citealt{Hirota_2017} and \citealt{JALV_2020}) and HH 30 (\citealt{Louvet_2018}). 
The apparent lack of molecular emission between $\sim$-1.5 km s$^{-1}$ and 1 km s$^{-1}$ observed in Figure \ref{fig:pvz}, 
could be due to the missing flux of the large--scale structure of the envelope. 

\begin{figure*}[t!]
\includegraphics[scale=0.55]{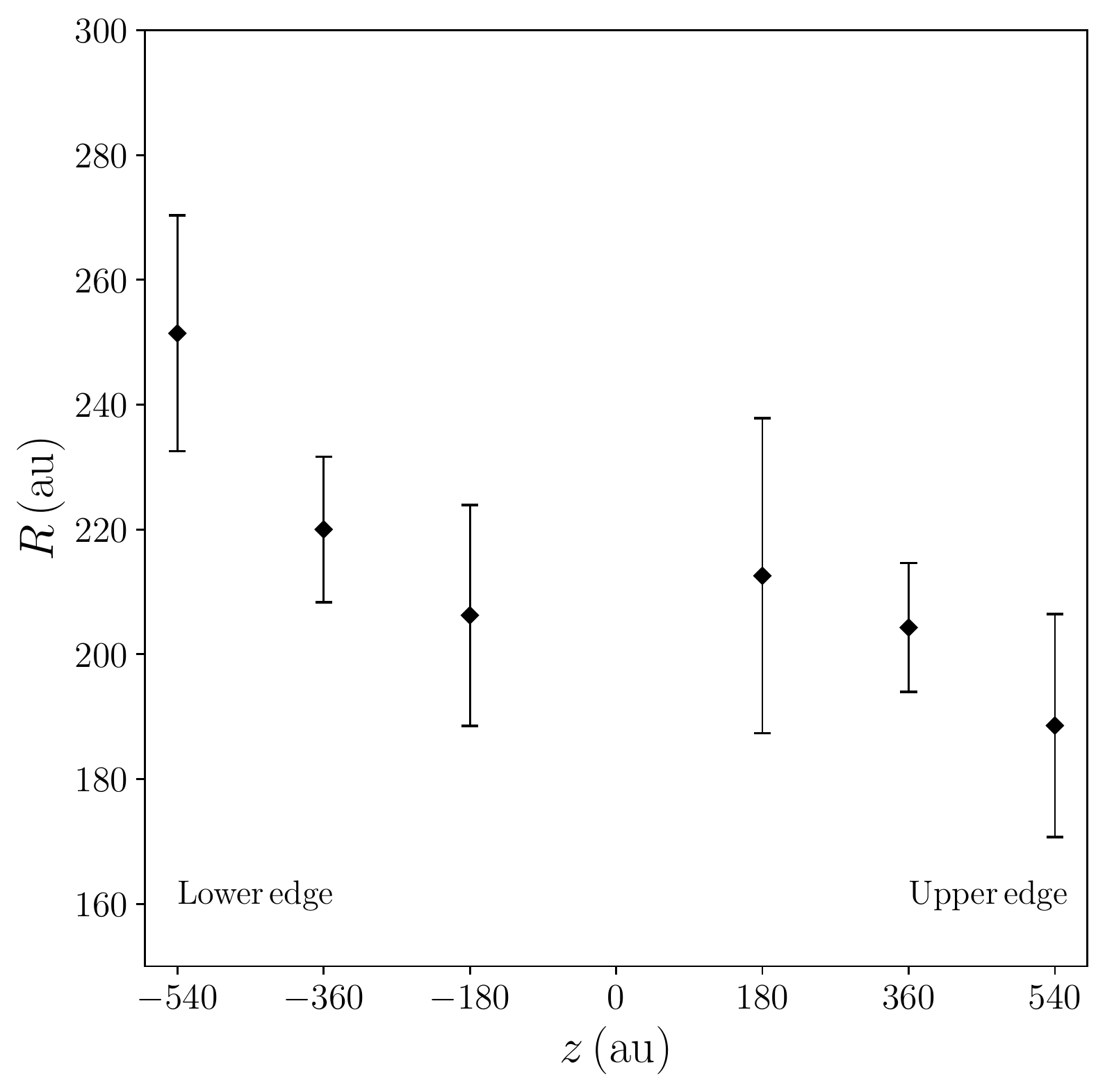}\includegraphics[scale=0.55]{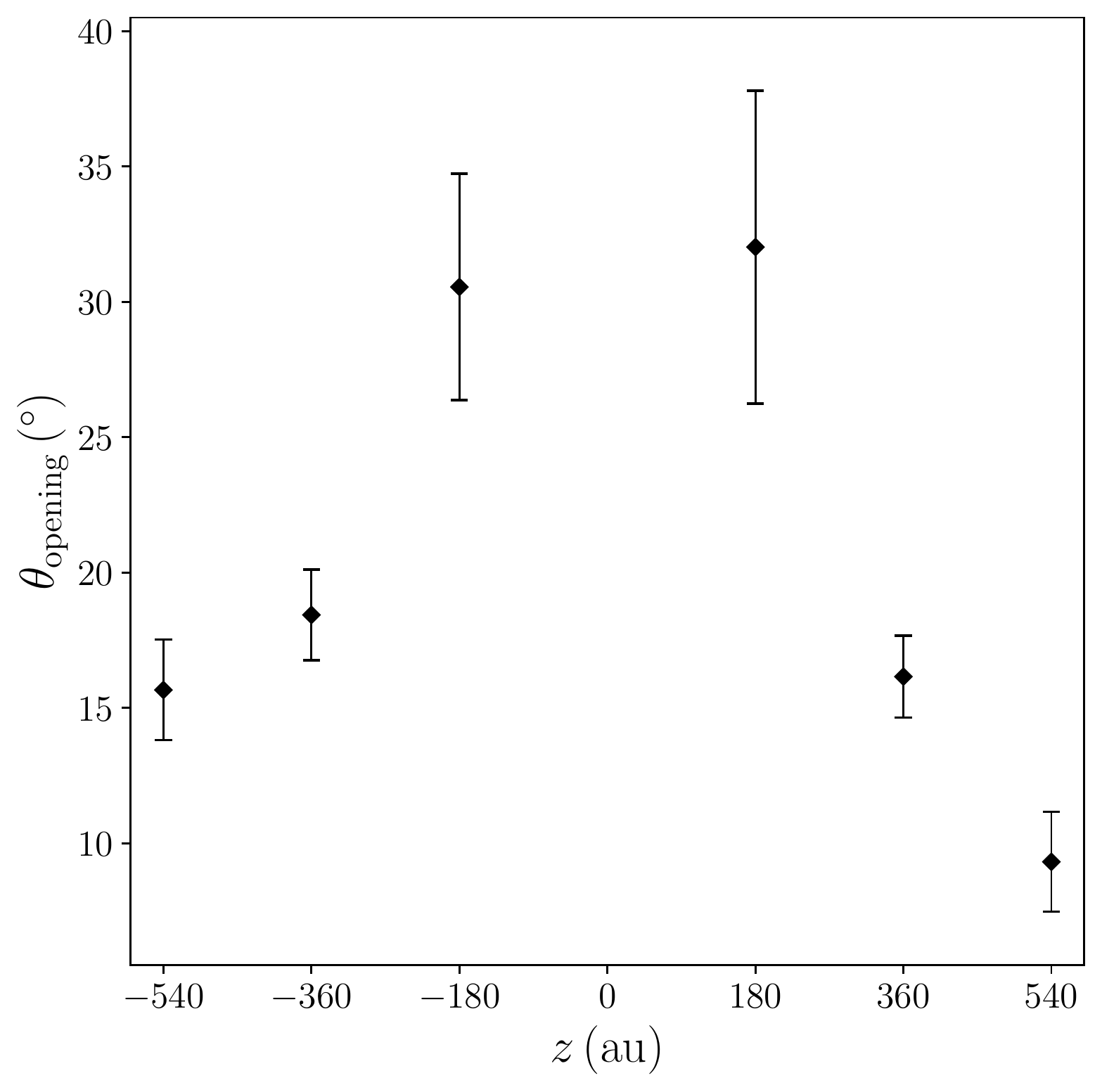}
\caption{Left panel: the radius of the outflow $R$. Right panel: the opening angle of the outflow $\theta_{\rm opening}$. 
These values are derived from the position-velocity diagrams in Figure \ref{fig:pvz}.
The error bars are derived from the gaussian fit.}
\label{fig:rang}
\end{figure*}

With the position--velocity diagrams presented in Figure \ref{fig:pvz}, we can obtain information about the kinematic and physical 
properties of the molecular outflow because they show rotation and expansion signatures as will be described later. These properties 
are: the radius $R$, the expansion velocity $v_{\rm exp}$, the rotation velocity $v_{rot}$, and additionally we obtain the opening 
angle $\theta_{\rm opening}$ and the specific angular momentum $j$. 
These quantities, except for the specific angular momentum, are measured by a Gaussian fit to the intensity profile as a function 
of the radius or velocity, as appropriate. For more details see appendix of \citet{JALV_2020} for the outflow in Orion Source I object. 
These measured quantities, the radius, the expansion, and the rotation velocities were not made for a height of $z=0$ au, 
this is because the accretion disk is located at this height.

The left panel of Figure \ref{fig:rang} shows the radius $R$ from the molecular outflow as a function of the height $z$. 
We note that these radii, in lower edge ($z<0$) of the molecular outflow, increase with the height above the disk mid-plane, however, 
this behavior is not evident in the upper edge $z>0$ from this outflow. The radius is in a range of $\sim$180--280 au, 
the error bars of these radii are obtained from the gaussian fit to the intensity profile as a function of the distance to the outflow axis. 
With these radii and  for fixed centrifugal radius, $R_{\rm cen}=200$ au, this radius corresponds to the disk radius measurement by \citet{Launhardt_2009},
the opening angle can be defined as (see Figure 6 of \citealt{JALV_2020})

\begin{eqnarray}
\theta_{\rm opening}=\tan^{-1}\left(\frac{R-R_{\rm cen}}{z}\right).
\label{eq:thetaop}
\end{eqnarray}

This angle is shown on the right panel of the Figure \ref{fig:rang}. \citet{Hirota_2017} and \citet{JALV_2020}, measured this angle 
for the molecular outflow associated with Orion Source I, and they found that this angle decreases with the height above the disk mid--plane. 
In this source, the opening angle also decreases with the height above the disk and presents the same behavior that reported by the molecular 
outflow of Orion Source I (\citealt{Hirota_2017}; \citealt{JALV_2020}). This angle is in a range of $\sim$9--32$^\circ$. 
The error bars are derived from the gaussian fit from the radii to the intensity profile. The fact that we used the centrifugal radius 
for our opening angle estimation is because this angle is a well indicator that the molecular outflow could close up at higher heights.

\begin{figure*}[t!]
\includegraphics[scale=0.55]{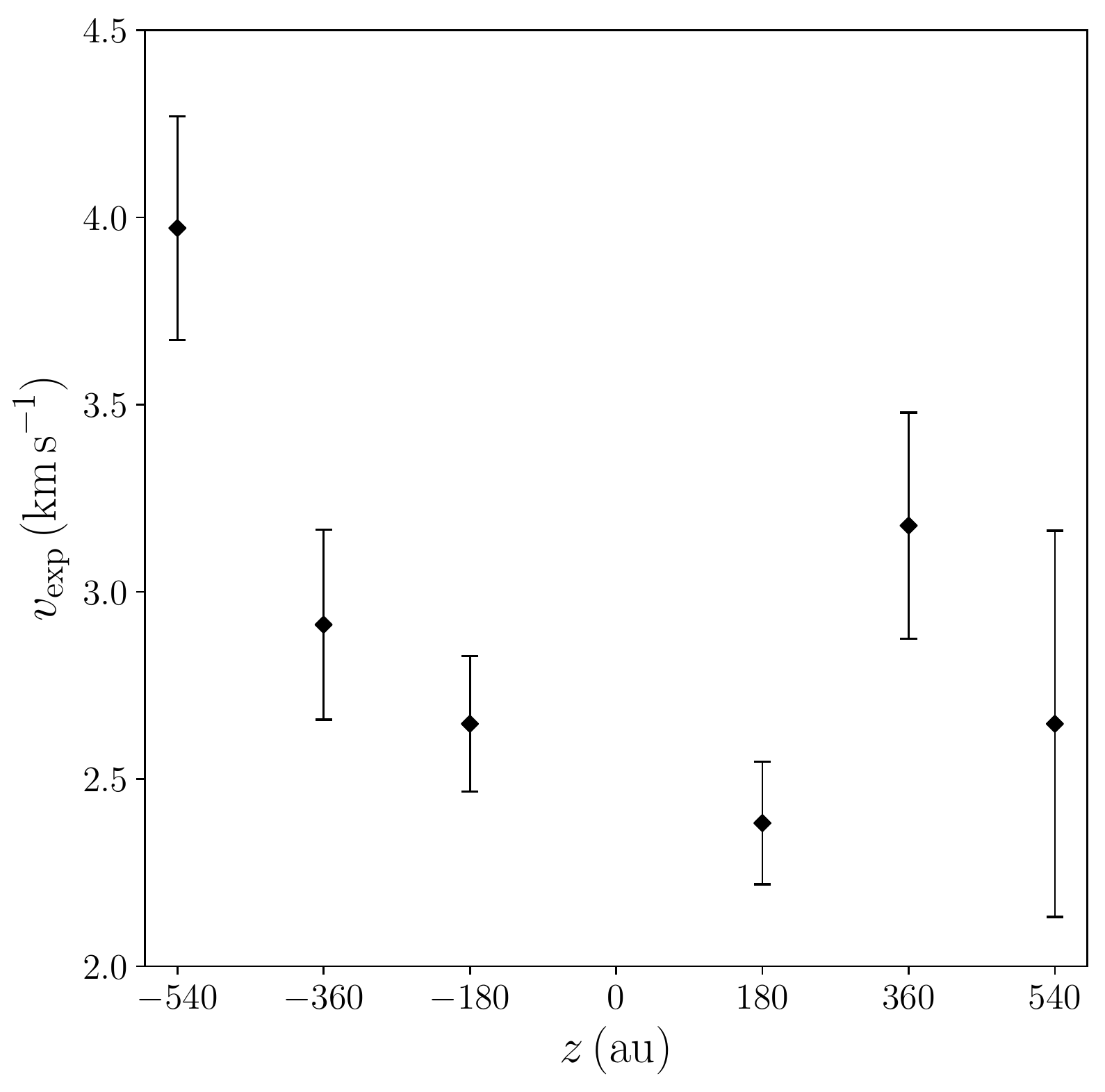}\includegraphics[scale=0.55]{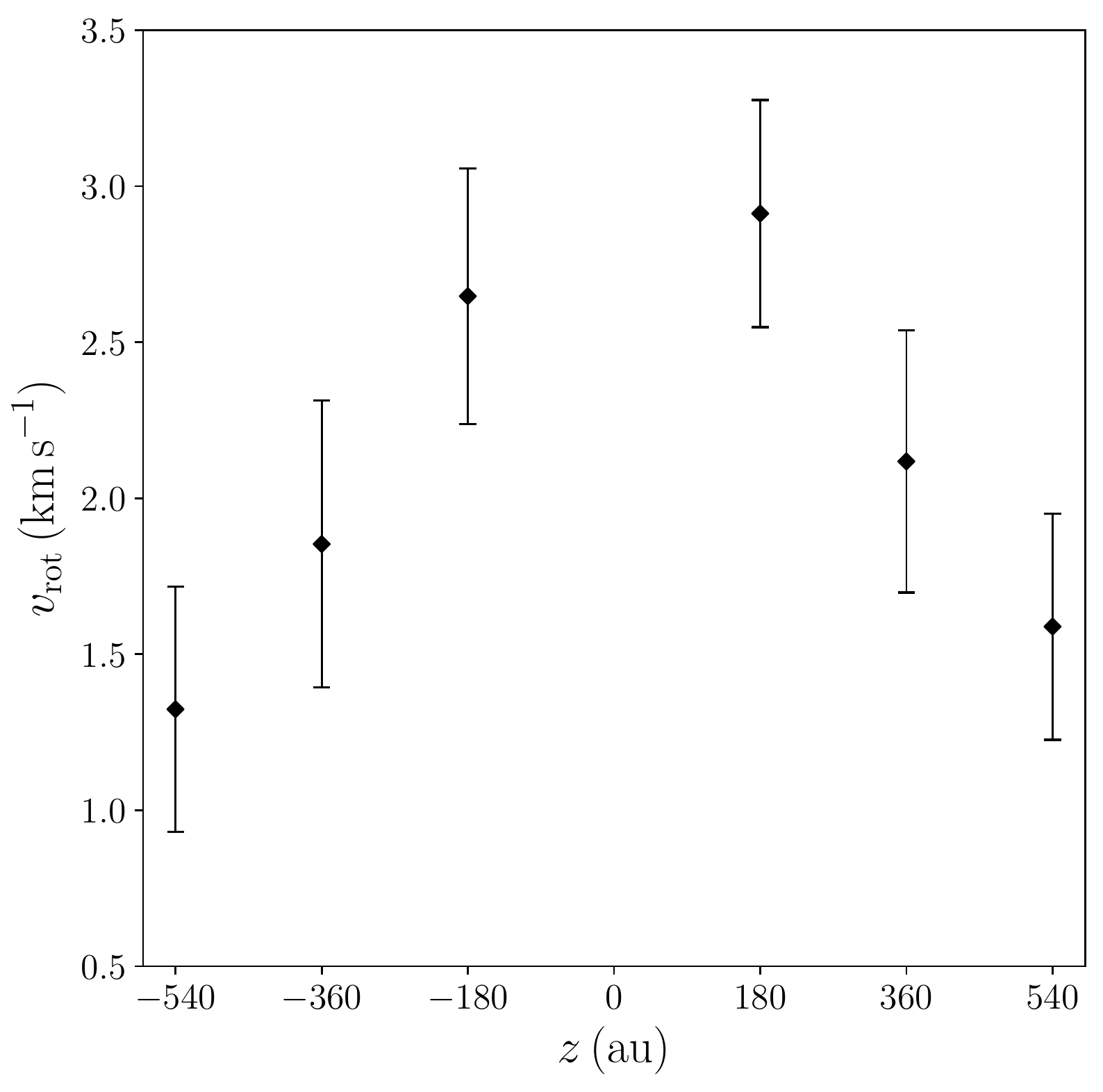}
\caption{Left panel: the expansion velocity perpendicular to the outflow axis $v_{\rm exp}$ measured at the radius $R$.
Right panel: the rotation velocity $v_{\rm rot}$ measured at the radius $R$.
These values are derived from the position-velocity diagrams  in Figures \ref{fig:pvz}.
The error bars are derived from the gaussian fit.}
\label{fig:vexprot}
\end{figure*}

\begin{figure}[t!]
\includegraphics[scale=0.55]{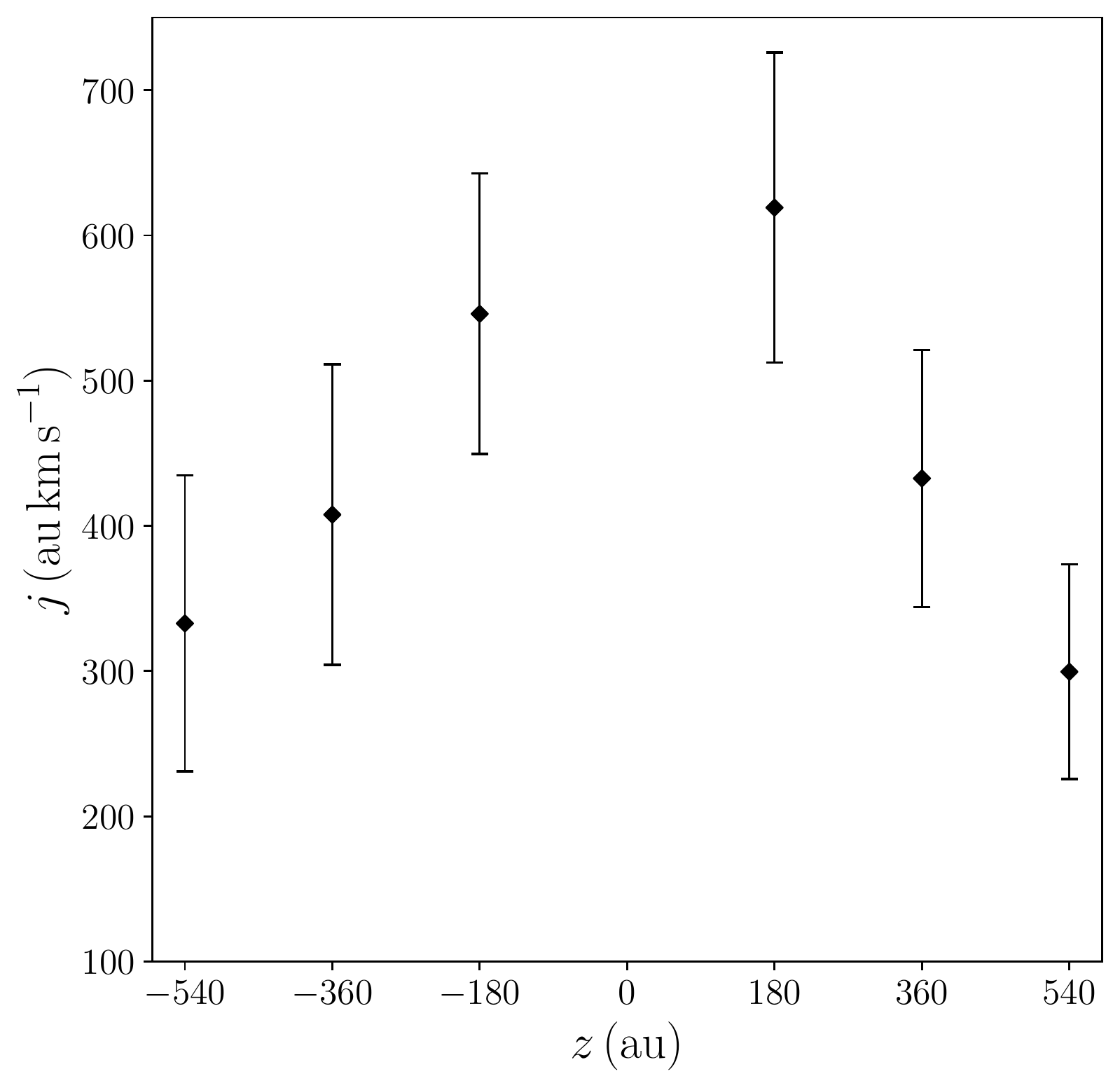}
\caption{The specific angular momentum $j$ as a function of the height $z$. The error bars are derived from the gaussian fit.}
\label{fig:specific}
\end{figure}

The left panel of Figure \ref{fig:vexprot} presents the projected expansion velocity $v_{\rm exp}$ of the line of sight
as a function of the height $z$. If we consider cylindrical coordinates, this velocity corresponds to the velocity in the radial direction (see \citealt{Tabone_2020}).
This velocity is in the range of 2--4 km s$^{-1}$ and increases with the height above the disk mid-plane, except for the height of $z$=540 au. 
Also, we can estimate the expansion velocity in the source system applying the correction $v_{\rm exp,source}\sim v_{\rm exp}/\cos i$, as the inclination 
angle respect to the plane of the sky is small ($\sim 5^\circ$), this results in an expansion velocity from the source system similar to the value of the projected expansion velocity.
While, the rotation velocity as a function of the height is shown in the right panel of Figure \ref{fig:vexprot}. 
This velocity is in the range of 1--3 km s$^{-1}$ and decreases with the height above the disk mid-plane. 
The error bars are obtained from the gaussian fit to the intensity profile as a function of the velocity (see Appendix of \citealt{JALV_2020}).

Figure \ref{fig:specific} present the specific angular momentum $j$ as a function of the height $z$. 
The specific angular momentum is calculated using the radius reported in left panel of Figure \ref{fig:rang} as $j=R v_{rot}$. 
The angular momentum decreases with the height. The specific angular momentum of the outflow is  $\sim$ 200--700 au km s$^{-1}$. 
The error bars are derived from the gaussian fits from the radii and rotation velocities to the intensity profile.

\subsection{Mass of the outflow, the disk, and the central source}
\label{subsec:mass}

Assuming local thermodynamic equilibrium and that the $^{12}$CO (J=2--1) molecular emission is optically 
thick, following the formalism of \citet{Zapata_2014}, the lower limit of the mass of the outflow is

\begin{eqnarray}
\left[\frac{M_{H_2}}{M_\odot}\right]&=&1.2\times10^{-15}X_{\frac{H_2}{CO}}\left[\frac{\Delta \Omega}{{\rm arcsec}^2}\right]\left[\frac{D}{\rm pc}\right]^2\nonumber \\
&\times& \left[\frac{{\rm exp}\left(\frac{5.53}{T_{ex}}\right)}{1-{\rm exp}\left(\frac{-11.06}{T_{ex}}\right)}\right]\left[\frac{\int I_\nu dv}{\rm Jy\,km\,s^{-1}}\right],
\label{eq:mass}
\end{eqnarray}

\noindent where $X_{\frac{\rm{H}_2}{\rm{CO}}}$ is the fractional abundance of $^{12}$CO with respect to H$_2$, for this source, 
we assumed an abundance of 7.5$\times10^{-5}$ \citep{Launhardt_2009}. The variable $\Delta \Omega$ is the solid angle of the
 source in arcsec$^2$, $D$ is the distance in parsec (140 pc), $I_\nu$ is the intensity of the emission of $^{12}$CO in Jy, $dv$ is 
 the velocity range in km s$^{-1}$, and $T_{\rm ex}$ is the excitation temperature given by {\citep{Estalella_1994}:
\begingroup
\footnotesize
\begin{eqnarray}
\left[\frac{T_{\rm ex}}{K}\right]=\frac{h\nu/k}{{\rm ln}\left(1+\frac{h\nu/k}{T_a+J_\nu\left(T_{bg}\right)}\right)}=\frac{11.07}{\mathrm{ln}\left(1+\frac{11.07 \rm{K}}{T_{\rm a}\left(^{12}{\rm CO}\right)+0.19 {\rm K}}\right)},
 \label{eq:tex}
\end{eqnarray}
\endgroup

\noindent  where $h$ is the Planck constant, $k$ is the Boltzmann constant, $\nu$ is
the rest frequency in GHz, $T_{\rm a}=16$ K is the observed antenna temperature given by the peak of the $^{12}$CO emission
and $J_\nu(T_{\rm bg})$ is intensity in units of temperature at the background temperature $T_{\rm bg}=2.7$ K. 
With this equation we obtain an excitation temperature of $T_{\rm ex} \approx  21$ K. Using this temperature and the values mentioned 
above, we estimate a mass for the outflow powered by CB 26 of $\rm{M}_{\rm outflow}\sim 5\pm1.5\times10^{-5}$ M$_\odot$. This value is lower 
by one order magnitude than the mass calculated by \citet{Launhardt_2009}. The difference between the mass obtained by us and the mass 
calculated by \citet{Launhardt_2009} could be because there are uncertainties in the column density $n_0$, and 
the kinetic gas temperature in the outflow $T_0$ (see section 4.2 of \citealt{Launhardt_2009}). 
The discrepancy found in the outflow mass by the current work and the mass calculated by \citet{Launhardt_2009}, 
could be due to the difference between the fluxes.

On the other hand, now assuming that the dust emission is optically thin, and using the relationship of \citet{Hildebrand_1983}, 
the dust mass of the protoplanetary disk is

\begin{figure}[t!]
\includegraphics[scale=0.52]{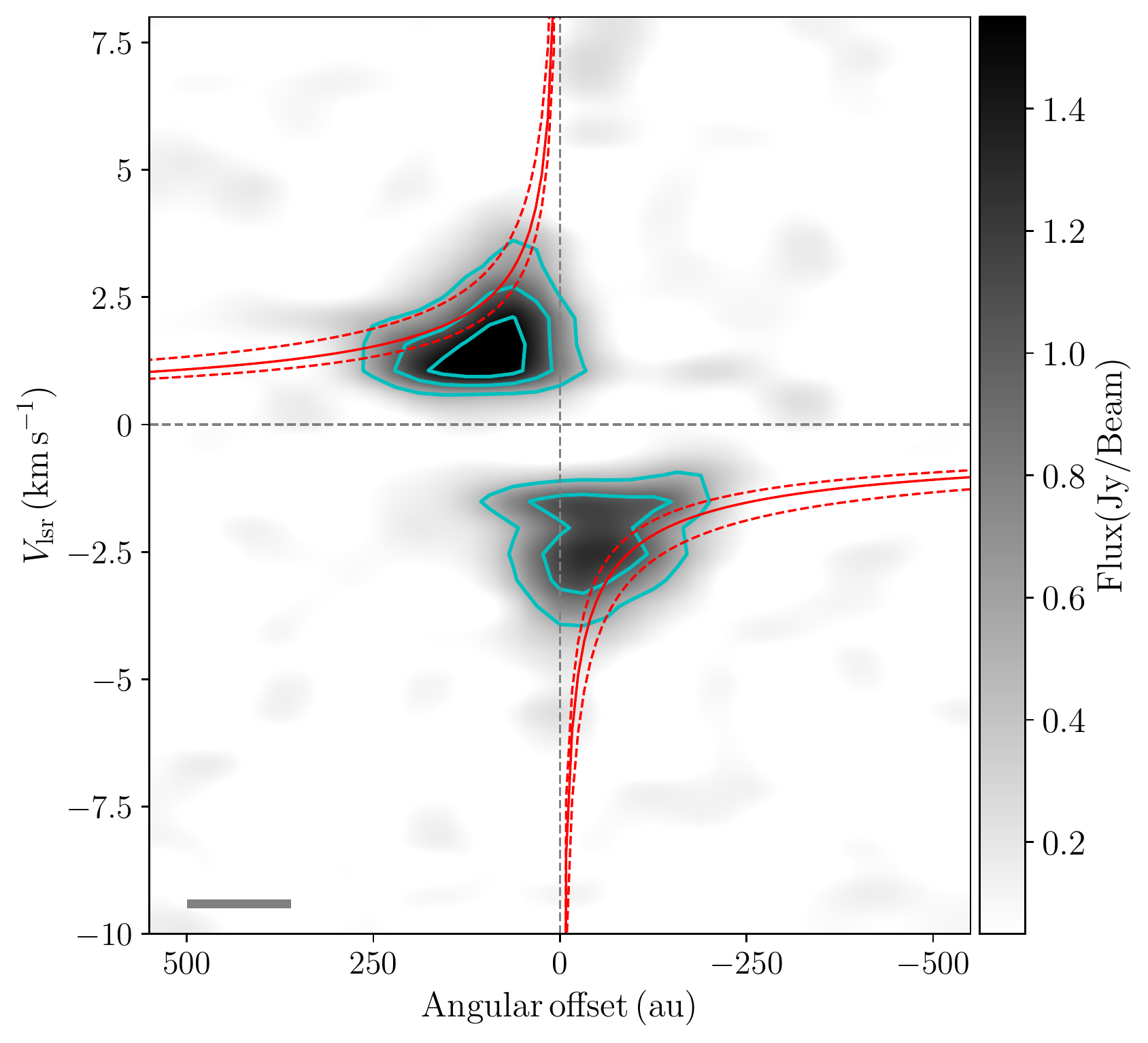}
\caption{Position--velocity diagram of $^{12}$CO (J=2--1) over the disk mid-plane ($z=$0). 
The color scale bar on the right side shows the intensity in Jy/beam. The red solid line shows the Keplerian velocity 
profile surrounding 0.66$\pm$0.03 M$_\odot$ central object.The gray bar represents the angular resolution (140 au or 1'') and velocity resolution (0.18 $\kms$) used for the position--velocity cut. Contours levels start from 5$\sigma$ in steps of 5$\sigma$, where 1$\sigma$ is 0.1012 Jy beam$^{-1}$.}
\label{fig:mdyn}
\end{figure}

\begingroup
\footnotesize
\begin{eqnarray}
M_{\rm d}=\frac{F_\nu D^2}{\kappa_\nu B_\nu\left(T_d\right)} \simeq 9.8\pm0.4\times10^{-7}\left(\frac{F_\nu}{\rm mJy}\right)\left(\frac{D}{100 {\rm pc}}\right)^2M_\odot,
\end{eqnarray}
\endgroup

\noindent  where $F_\nu$ is the mm flux, $D$ is the distance to the source, $T_d$ is the dust temperature, $B_\nu$ is the Planck function at $T_d$, 
and $\kappa_\nu$ is the dust grain opacity. We use standard assumptions for the values of the parameters $T_d=20$ K (\citealt{Andrews_2005}),
 and a power--law opacity of the form $\kappa_\nu=10 (\nu/1000\,{\rm GHz})^\beta$ cm$^2$g$^{-1}$, with a $\beta$ value of 1.1$\pm$0.27 (\citealt{Sauter_2009}). 
 Using the above equation, we found a dust mass of $M_{\rm dust}=$ 0.00031$\pm$0.00015 M$_\odot$, so taking a typical ratio between the gas and dust in the ISM of 100,
 we obtain a total mass for the disk of 0.031$\pm$0.015 M$_\odot$. 
 
In order to estimate the dynamical mass of the central object, we fit a Keplerian velocity profile to the position--velocity diagram of the disk ($z$=0 au), 
this fit is shown in Figure \ref{fig:mdyn}. The best fit (red line), corresponds to a dynamical mass of M$_{dyn}$=0.66$\pm$0.03 M$_\odot$ while the dashed lines
are 1.0 (outer lines) and 0.5 M$_\odot$ (inner lines), the best fit corresponds to the visual fit of a Keplerian curve $v_k=\sqrt{G M_{\rm dyn}/r}$ where $v_K$ and $r$ are measured at the contours corresponding to the emission of 10$\sigma$ of Figure \ref{fig:mdyn}.

\subsection{Origin of the molecular outflow}
\label{subsec:origin}

\begin{figure}[t!]
\includegraphics[scale=0.55]{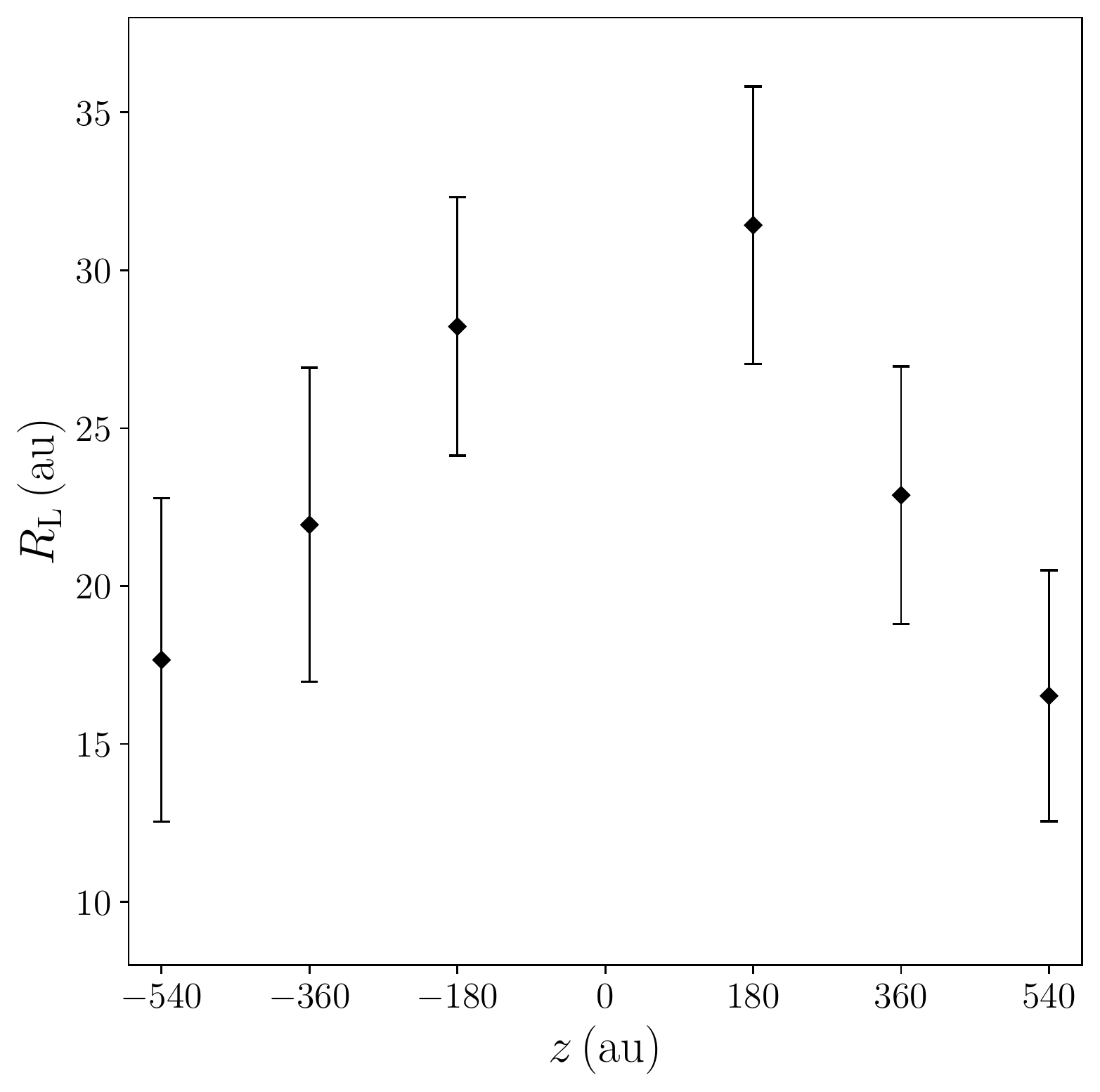}
\caption{The launching radii $R_L$ as a function of the height $z$. These radii are calculated solving the eq. (4) of \citet{Anderson_2003} (see text). 
The error bars are derived from the gaussian fit.}
\label{fig:rlaunch}
\end{figure}

As mentioned in section \ref{sec:introduction}, the origin of the molecular outflow is uncertain. A possibility is that the wind is launched 
through magneto-centrifugal processes from a rotating protostellar disk and then accelerated and collimated by magnetohydrodynamic forces. 
Under this scenario, we can calculate the launching radius following the equation (4) of \citet{Anderson_2003} given by
\begin{eqnarray}
\varpi_\infty v_{\phi, \infty}\Omega_0-\frac{3}{2}\left(\rm GM_*\right)^{2/3}\Omega_0^{2/3}-\frac{v_{\rm p,\infty}^{2}+v_{\phi,\infty}^2}{2}\approx0,
\label{eq:rlaunch}
\end{eqnarray}

\noindent where $\varpi_\infty$ is the observed radial distance to the flow axis, in our case is $R$ for each $z$, $v_{\phi, \infty}$ and $v_{\rm p,\infty}$ 
are the toroidal and poloidal velocities observed at radius $R$. For this object, the toroidal velocity corresponds to the rotation velocity $v_{\rm rot}$, 
while the outward velocity of the molecular outflow is $v_{\rm r}\sim$10 km s$^{-1}$ \citep{Launhardt_2009}, the outflow inclination angle is $i=5^\circ\pm4^\circ$, 
thus, the corrected outward velocity is $v_{\rm z}=v_{\rm r}/\cos i\sim 10.04\pm0.06$ km s$^{-1}$, and the poloidal velocity is $v_p=\sqrt{v^2_z+v^2_{\rm exp}}$. The gravitational constant is G 
and the mass of the central protostar is M$_*$, we assume  M$_*$=0.6 M$_\odot$, a similar value to that estimated in \citet{Zhang_2021}. 
Finally, $\Omega_0$ is the angular speed at the launching radius, given by

\begin{eqnarray}
\Omega_0=\left(\frac{\rm GM_*}{\varpi_0^3}\right)^{1/2},
\label{eq:aspeed}
\end{eqnarray}

\noindent where $\varpi_0$ is the launching radius $R_{\rm L}$. 

Solving Eq. (\ref{eq:rlaunch}) for $R_L$ with the values mentioned above, we calculate the launching radius of the molecular outflow, 
this radius is shown as a function of the height above the disk mid-plane in Figure \ref{fig:rlaunch}. We note that these radii (in a range of $\sim$ 15--35 au) 
decreases with the height.

\section{Discussion}
\label{sec:discussion}

\begin{table*}[!t]
 \centering
 \setlength\tabcolsep{2.0pt}
\caption{Derived quantities of the molecular outflow and disk.}
\begin{tabular}{l l l}
 \hline
 \hline
\textbf{Parameter} & \textbf{Symbol} & \textbf{Value }   \\
\hline
Outflow mass & M$_{\rm outflow}$ & 5$\pm$1.5$\times$10$^{-5}$ M$_\odot$\\
Outflow mass loss rate & $\dot{\rm M}_{\rm outflow}$ & 2$\pm$0.6$\times$10$^{-7}$ M$_\odot$yr$^{-1}$\\
Outflow linear momentum rate &$\dot{P}_{\rm outflow}$ & 2$\pm$0.7$\times$10$^{-6}$ M$_\odot$yr$^{-1}$ km s$^{-1}$\\
Outflow angular momentum rate & $\dot{L}_{\rm outflow}$ & 6$\pm$2.4$\times$10$^{-5}$ M$_\odot$ yr$^{-1}$ au km s$^{-1}$ \\
Disk mass & M$_d$ & 0.031$\pm$0.015 M$_\odot$ \\
Dynamical mass (central object) & M$_{dyn}$ & 0.66$\pm$0.03 M$_\odot$\\
\hline
\end{tabular}
  \label{tab:quantities}
\end{table*}

Figures \ref{fig:rang}--\ref{fig:specific} show the structure and kinematic of the molecular outflow associated with CB 26. One can note that the radius, 
the opening angle, and the rotation velocity have a similar behavior to other rotating molecular outflows, {\it e.g.}, Orion Source I (\citealt{Hirota_2017} and \citealt{JALV_2020}), 
HH 30 \citep{Louvet_2018}, and NGC 1333 IRAS 4C \citep{Zhang_2018}, this means that the radius increase with the height above the disk mid plane, 
and the opening angle and the rotation velocity decreases with the height above the disk mid-plane. With respect to the specific angular momentum, it seems to have 
the same behavior to the sources mentioned above, this is, it decreases with the height. In summary, a disk-wind driving the outflow could explain all the 
observed characteristics. 

The masses estimated for the molecular outflow, $5\pm1.5\times10^{-5}$ M$_\odot$, the protostellar disk, $0.06\pm 0.017$, and the central star, $\sim 0.6\pm0.03$ M$_\odot$, 
are all consistent with the values of the literature (e.g., \citealt{Launhardt_2009}; \citealt{Zhang_2021}).

In section \ref{subsec:origin} we estimated the launching radius of the wind under assumption that the origin of the molecular outflow is the large radius 
of the Keplerian disk. This wind can be ejected by magneto-centrifugal mechanism or by photoevaporated disk wind. Nevertheless, \citet{Zapata_2015} 
showed that magnetocentrifugal and photoevaporated disk winds do not have enough mass to account for the observed mass rates in the molecular outflow
of DG Tau B, their argument is based on the assumption that the mass-loss rate of the wind is a fraction $f\sim 0.1$ of the mass accretion rate, $\dot{M}_w\sim f\dot{M}_{\rm d,a}$.
However, we note that recent ALMA observations for DG Tau B \citep{DeValon_2020} reported a lower outflow mass (a factor of 30), 
and suggested that a disk-wind could be responsible for the rotating outflow, however it still being massive to explain the outflow from DG Tau B.
 
In particular this source has an outflow mass of 5$\times10^{-5}$ M$_\odot$ (see section \ref{subsec:mass}), for a corrected outward velocity $v_{\rm z}\sim$ 10.04$\pm$0.06 km s$^{-1}$ 
and a projected size (diameter) of $z\sim$ 540 au, the kinematic time is $t_{\rm kin}\sim$ 255$\pm1.6$ yr. Then, the molecular outflow mass loss rate is 
$\dot{M}_{\rm outflow}=M_{\rm outflow}/t_{\rm kin}\sim 2\pm0.012\times10^{-7}\,M_\odot$ yr$^{-1}$, its linear momentum rate 
$\dot{P}_{\rm outflow}=\dot{M}_{\rm outflow} v_{\rm p}\sim 2\pm 0.035\times10^{-6}\,M_\odot$ yr$^{-1}$ km s$^{-1}$ or a height of $z=540$ au its angular momentum rate is 
$\dot{L}_{\rm outflow}=\dot{M}_{\rm outflow }R v_{\rm rot}\sim 6 \pm 1.5 \times 10^{-5}\,M_\odot$ yr$^{-1}$ au km s$^{-1}$ $=9\pm 2.4\times10^3\,M_\odot$ yr$^{-1}$ km$^2$ s$^{-1}$.  
These estimated rates, as well as the outflow mass, the disk mass, and the dynamical mass of the central object are shown in Table \ref{tab:quantities}.

If we take a value for the mass-loss rate for a disk-wind of $\dot{M}_{\rm wind}$ $\sim$ 10$^{-7}$ M$_\odot$ yr$^{-1}$, 
which is a higher value than that reported in DG Tau B, a Class I/II object, see \citet{Zapata_2015}. The expected mass-loss rate is thought to increase in younger objects, 
so this assumption should be reasonable. Thus, the linear momentum rate for the slow disk-wind is $\dot P_w = \dot M_w \, v_p \sim  1 \pm 0.014 \times 10^{-6} \msun \, \yr \, \kms $, and the
 angular momentum rate is  $\dot L_w = \dot M_w r v_\phi = 3  \pm 0.7 \times 10^{-5} \msun\,\yr$ au $\kms=4.5\pm 1.1 \times 10^{3}\msun\,\yr\,{\rm km^2}\,{\rm s}^{-1}$. Both rates
are very similar to the ones obtained to the molecular outflow in CB26 ($\dot{P}_{\rm outflow}$ and $\dot{L}_{\rm outflow}$), which indicates that a disk-wind has enough momentum to 
drive the molecular outflow.

If the outflow is a disk wind, $\dot{M}_{\rm outflow}=\dot{M}_w\sim f\dot{M}_{\rm d,a}$, we can estimate the accretion luminosity at the stellar surface as

\begin{eqnarray}
L_a=\eta \frac{G M_*\dot{M}_{\rm d,a}}{R_*},
\end{eqnarray}

\noindent where $G$ is the gravitational constant (as above), $M_*$ is the stellar mass, $R_*$ is the stellar radius, 
and $\eta\sim0.5$. Assuming $M_*=0.6$ M$_\odot$ and $R_*=2$ R$_{\odot}$ (\citealt{Zhang_2021}), 
the accretion luminosity is $L_a\geq (1/f)0.9$ L$_\odot$. 
This value is consistent to the reported value of the central source $L_*=0.92$ L$_\odot$ (\citealt{Zhang_2021}) by a factor of $f\sim1$.

In summary,  all the presented observational characteristics argue in favor of having a disk-wind in CB26 outflow.  The disk-wind for this case has enough mass to account to the mass rates.
 
\section{Conclusions}
\label{sec:Conclusions}

In this work, we present the SMA archive observations for the line emission of $^{12}$CO (J=2--1)  from the molecular outflow CB 26. Our main results are:

\begin{enumerate}
\item We find that the radius $R$ (in a range of $\sim$ 180--280 au) and the expansion velocity (in a range of 2--4 km s$^{-1}$) increase with the 
height above the disk mid--plane in lower edge of the molecular outflow ($z<$0 au), however, this behavior is not evident in the upper edge ($z>$0 au).

\item We find that the rotation velocity (in a range of 1--3 km s$^{-1}$), the specific angular momentum (in a range of 200--700 au km s$^{-1}$),
 and the launching radius (in a range of 15--35 au),  decrease with the height above the disk mid--plane, as observed in other molecular rotating outflows.
 
\item We estimated the mass of the molecular outflow, $5\pm1.5\times10^{-5}$ M$_\odot$, the mass of the accretion disk, $0.031\pm0.015$ M$_\odot$, and the mass of the central star $0.66\pm0.03$ M$_\odot$.

\item Estimations of the outflow linear momentum rate, the outflow angular momentum rate, and the accretion luminosity seem to be well explained 
by a disk-wind present in CB26. 

\end{enumerate}




\listofchanges

\end{document}